\documentclass[pra,twocolumn,titlepage,nofootinbib,amsmath]{revtex4}
\usepackage{amsmath}
\usepackage{amsfonts}
\usepackage{amssymb}
\usepackage{graphicx}
\begin{document}
\title{Relativistic recoil corrections to the electron-vacuum-polarization
contribution\\ in light muonic atoms}
\author{Savely~G.~Karshenboim}
\email{savely.karshenboim@mpq.mpg.de}
\affiliation{Max-Planck-Institut f\"ur Quantenoptik, Garching,
85748, Germany} \affiliation{Pulkovo Observatory, St.Petersburg,
196140, Russia}
\author{Vladimir~G.~Ivanov}
\affiliation{Pulkovo Observatory, St.Petersburg, 196140, Russia}
\author{Evgeny~Yu.~Korzinin}
\affiliation{D.~I. Mendeleev Institute for Metrology, St.Petersburg,
190005, Russia}

\begin{abstract}
The relativistic recoil contributions to the Uehling corrections are
revisited. We consider a controversy in recent calculations,
which are based on different approaches including
Breit-type and  Grotch-type calculations.

We have found that calculations of those authors were in fact done
in different gauges and in some of those gauges contributions to
retardation and two-photon-exchange effects were missed. We have
evaluated such effects and obtained a consistent result.

We present a correct expression for the Grotch-type approach which
produces a correct gauge-invariant result.

We also consider a finite-nuclear-size correction for the Uehling
term.

The results are presented for muonic hydrogen and deuterium atoms and for muonic helium-3 and helium-4 ions.
\end{abstract}

\maketitle

\section{Introduction}

A recent experiment performed at PSI on muonic-hydrogen Lamb
shift \cite{nature} has reported a high-precision
result on the proton charge radius. This result is in a strong
contradiction with a recent electron-proton scattering result from
MAMI~\cite{mainz} and the CODATA-2006 value~\cite{codata}, which
basically originates from the hydrogen and deuterium spectroscopy and
involves large amount of experimental data and theoretical
calculations.

The discrepancy is at the level of 0.3~meV in terms of the
muonic-hydrogen Lamb shift. Meanwhile the theoretical uncertainty is
equal to 0.004~meV and that from experiment is 0.003~meV.
It is highly unlikely that the problem lies in either
theory or experiment on muonic hydrogen. Nevertheless, it is
important to clarify the theory of the muonic-hydrogen Lamb shift.

We expect that the controversy will be resolved and the
muonic-hydrogen Lamb shift will become the most accurate way to
determine a value of the proton charge radius. For this reason it is
important to have a reliable theoretical expression at the level of
0.003~meV~\cite{nature}.

The theoretical expression consists of quantum-electrodynamics
contributions and finite-nuclear-size corrections.

After a calculation of all the light-by-light contributions in order
$\alpha^5m_\mu$~\cite{LbL,LbL2} the quantum electrodynamics theory at
this level of uncertainty is complete in a sense that all
corrections have been calculated by at least one author or one
group.

However, verification is required and we consider certain
corrections as not well established. That in particular includes a
contribution of the recoil effects in order $\alpha(Z\alpha)^4m_\mu$
due to electronic-vacuum-polarization effects.

The electronic-vacuum-polarization (eVP) effects form one of the
most important features of a muonic atom which distinguish it from
an ordinary atom. The leading eVP effect is due to a so-called
Uehling potential and it produces the leading contribution to the
Lamb shift in light muonic atoms. The correction is of order
$\alpha(Z\alpha)^2m_\mu$. Since that is the largest contribution, it is
important to calculate the eVP terms including various higher-order
effects.

Another important feature is that the ratio of the mass $m$ of the
orbiting particle, a muon, is smaller than the nuclear mass $M$, but
not so small as in an conventional atom and, in particular,
$m/M\simeq 1/2000$ in ordinary hydrogen, while $m/M\simeq 1/9$ in
muonic hydrogen. That is why we need to find recoil corrections
for  most contributions of interest.

The purpose of this work is to obtain an $\alpha(Z\alpha)^4m_\mu$
contribution in all orders of $m/M$ in light muonic atoms ($Z=1,2$).
The most straightforward way is to derive a non-relativistic
expansion for both particles within a Breit-type
equation~\cite{breit,breit1}. However, for comparison we also use a
Grotch-type technique~\cite{grotch}. The latter produces a
non-recoil term and the $m/M$ leading recoil correction, while the
$(m/M)^2$ term is to be calculated separately. Both methods are
considered in detail in this paper.

These two methods are quite different. The Breit-type evaluation
involves a non-relativistic expansion and the related perturbation
theory includes terms of first and second order. Most of
the first-order terms appear naturally in momentum space, which is
more challenging for a high-accuracy calculation.

The Grotch-type approach involves non-trivial analytic
transformations and as a result the expression for most of the
energy contributions is obtained analytically in terms of a solution
of the ordinary Dirac equation with a static potential. The only
additional correction is expressed as a first-order perturbation,
which for our purposes may be calculated using non-relativistic
Coulomb wave functions.

The results originally published for the relativistic recoil
contribution within these different
approaches~\cite{pachucki1,borie,jentschura} are not consistent. The size of
discrepancy is comparable to the experimental uncertainty~\cite{nature}.

Here, we find that in fact calculations in different approaches are
inconsistent because different gauges were used and in one of the
calculations certain additional terms should appear. Those terms
originate from retardation effects and from essential two-photon
contributions and after proper corrections we find that the two
approaches are consistent and produce the same result for the
$\alpha(Z\alpha)^4m(m/M)$ terms.

Comparing different approaches and comparing our results with the
results of other authors we basically focus our attention on muonic
hydrogen. In addition, in summary sections we also discuss other
$Z=1,2$ two-body muonic atoms. Units in which $\hbar=c=1$ are
adopted throughout the paper.

The paper is organized as follows. We first consider the
non-relativistic leading eVP contribution and the
leading relativistic term in order $\alpha(Z\alpha)^4m$. Then
we discuss different gauges to take into account eVP effects. The two
gauges we choose are closely related to static potentials applied
in~\cite{pachucki1,jentschura,borie}. Both gauges are defined as
a certain modification of the Coulomb gauge.

We apply the Breit-type and Grotch-type approaches to calculate
relativistic recoil contributions in the gauges and obtain consistent
results in order $\alpha(Z\alpha)^4m(m/M)$. We find additional
contributions due to a one-photon retardation contribution and due to
two-photon exchange effects in one of those gauges. We demonstrate
that the effective potentials generated by those additional terms
agree with the difference in static potentials in the two gauges.

We consider the term of order $\alpha(Z\alpha)^4m(m/M)^2$ only in
the Breit-type approach. Since the value of this contribution depends on
the definition of the nuclear radius, which may `absorb' part of the
correction, we recalculate a nuclear-finite-size term and obtain a
semi-analytic result for it.

\section{The leading non-relativistic and relativistic ${\rm\textbf{eVP}}$ terms
and the relativistic recoil effects}
\label{s:leading}

The non-relativistic (NR) Uehling term can be easily calculated both
analytically~\cite{uehl_cl,uehl_an} and numerically (see, e.g.,~\cite{uehl_num})
for an arbitrary hydrogenic state (with the reduced-mass corrections included).
In particular, the results for the low states in
light muonic atoms are~\cite{uehl_cl}
\begin{eqnarray}\label{u:lead}
E^{(\rm NR)}_{\rm VP}(nl)&=&\frac{\alpha}{\pi}\frac{(Z\alpha)^2m_R}{n^2}\,F_{nl}(\kappa/n)\;,\\
F_{1s}(z)&=&- \frac{1}{3}\biggl\{-\frac{4+z^2-2\,z^4}{z^3}\cdot A(z)\nonumber\\
&~&+\frac{4+3\,z^2}{z^3}\cdot \frac{\pi}{2}-\frac{12+11\,z^2}{3\,z^2} \biggr\}\;,\nonumber\\
F_{2s}(z)&=&- \frac{2}{3}\biggl\{-\frac{16}{3}-\frac{14}{z^2}+\pi\left(\frac{3}{2z}+\frac{7}{z^3}\right)\nonumber\\
&~&+\frac{3}{4}\frac{z^2}{z^2-1}+\frac{9}{4}\frac{z^4}{(z^2-1)^2}\nonumber\\
&~&+z A(z)\left[\frac{13}{4}+\frac{4}{z^2}-\frac{14}{z^4}-\frac{9}{4}\frac{z^4}{(z^2-1)^2}\right] \biggr\}\;,\nonumber\\
F_{2p}(z)&=&- \frac{2}{3}\biggl\{-\frac{14}{3}-\frac{10}{z^2}+\pi\left(\frac{3}{2z}+\frac{5}{z^3}\right)\nonumber\\
&~&+\frac{5}{4}\frac{z^2}{z^2-1}+\frac{3}{4}\frac{z^4}{(z^2-1)^2}+z A(z)\nonumber\\
&~&\times\left[\frac{11}{4}+\frac{2}{z^2}-\frac{10}{z^4}-\frac{z^2}{z^2-1}-\frac{3}{4}\frac{z^4}{(z^2-1)^2}\right]
\biggr\}\;,\nonumber
\end{eqnarray}
where
\begin{equation}\label{Adef}
  A(z)=\frac{\arccos(z)}{\sqrt{1-z^2}}=
  \frac{\ln\left(z+\sqrt{z^2-1}\right)}{\sqrt{z^2-1}}\,,\\
\end{equation}
\begin{equation}
\kappa=Z\alpha m_R/m_e\;,
\end{equation}
$m_R$ is the reduced mass
\[m_R=\frac{mM}{M+m}\;,\]
$m$ and $M$ are masses of the muon and nucleus respectively. We remind that in muonic hydrogen
$\kappa\simeq 1.5$ and a characteristic value for the $2s$ and $2p$
states is $\kappa/2\simeq 0.75$. In muonic helium ion this value is
$\kappa/2\simeq 1.5$.

It is convenient to present the relativistic eVP correction in order
$\alpha(Z\alpha)^4m$ as an expansion in powers of $m/M$
\begin{eqnarray}\label{recoilsplit}
E_{\rm VP}^{\rm(rel)}&=&E_{\rm VP}^{(0)} +E_{\rm VP}^{(1)}
+E_{\rm VP}^{\rm(2)}+\dots\nonumber\\
&=&\frac{\alpha}{\pi}\,(Z\alpha)^4m_R\biggl[C_0
(\kappa)+C_1 (\kappa)\,\frac{m}{M}\nonumber\\
&~&+C_2 (\kappa)\,\left(\frac{m}{M}\right)^2+\dots \biggr]
\end{eqnarray}

In this notation $E_{\rm VP}^{(0)}$ and $C_0 (\kappa)$ are related to the
leading relativistic correction, which, e.g., may be obtained from
the Dirac equation including the Coulomb and Uehling potentials for a
muon with the reduced mass. Such relativistic non-recoil
contributions can be found through the Dirac-Uehling equation both
semi-analytically~\cite{rel1,rel2} and
numerically~\cite{uehl_num,pachucki2,borie} for various states.

In particular, for the $n=1,2$ states in muonic hydrogen
($\kappa=1.356146\dots$) we find from~\cite{rel2}
\begin{eqnarray}\label{allc0}
C_0(1s)&=& -0.24488\dots\;. \nonumber\\
C_0(2s)&=& -0.042224\dots\;. \nonumber\\
C_0(2p_{1/2})&=&-0.0089077\dots\;. \nonumber\\
C_0(2p_{3/2})&=& -0.00089011\dots\;,
\end{eqnarray}
which indeed agree with the results
from~\cite{uehl_num,pachucki2,borie,jentschura}.

The next coefficient, $C_1$, can be obtained in quite different
approaches (cf.~\cite{pachucki1,borie,jentschura}) and one has to be
careful while classifying contributions by counting photon
exchanges. A rigorous consideration could start with a
Bethe-Salpeter equation and, through rearranging its kernel, arrive
at an effective one-particle or two-particle equation. When we refer
here to one-photon exchange, we mean that the kernel of such an equation
contains only a one-photon part.

Any solution of an unperturbed problem is a summation over an
infinite number of such one-photon exchanges. But all those
many-photon exchange diagrams are reducible.

The approach based on a Grotch-type equation (see, e.g.,~\cite{borie})
immediately produces an appropriate summation for the
pure Coulomb exchange for the $(Z\alpha)^4m$ and $(Z\alpha)^4m^2/M$
terms. Meanwhile to take into account the eVP contribution one must use
a perturbation theory, but only in its leading order.

When the same physical contributions are treated
non-relativistically, e.g., by applying the Breit-type approach
(see \cite{pachucki1,jentschura}),
one has to use a certain effective $(v/c)^2$ expansion.
A perturbation theory is to be introduced already for a pure Coulomb
problem to reach $(Z\alpha)^4m$ (both for the leading non-recoil
term and for recoil corrections). As a result, a part of the relativistic
one-photon contributions for $\alpha(Z\alpha)^4m$ turns into
reducible two-photon exchange contributions.

One should distinguish between such reducible two-photon
contributions and irreducible two-photon diagrams which
appear while calculating $(Z\alpha)^4m^2/M$ in, e.g., covariant
gauges. It happens that neglecting such irreducible contributions in one of
the former calculations, namely in~\cite{borie}, produces a result which
is incomplete and must be corrected as considered below.

The leading relativistic-recoil correction beyond the Dirac-Uehling
term is of order $\alpha(Z\alpha)^4m^2/M$ and may be calculated, as
we mentioned, via various effective two-body techniques.

The $\alpha(Z\alpha)^4m$ term including various recoil corrections
was calculated previously by a number of authors. Their results for
$E_{\rm VP}^{\rm(rel)}(2p-2s)$ are 0.0169~meV~\cite{pachucki1},
0.0169~meV~\cite{borie} and 0.018759~meV~\cite{jentschura} and look
consistent at first glance.

However, as we note, the Dirac term
\begin{equation}
E_{\rm VP}^{(0)}(2p_{1/2}-2s_{1/2})=0.020843\dots\,{\rm meV}
\end{equation}
has never been a problem and had been known from calculation of
various authors for a while before the results mentioned above were
achieved (see, e.g. \cite{pachucki2,uehl_num}). Once we subtract the
Dirac term the results become strongly contradicting and the
discrepancy is compatible with the uncertainty of 0.003~meV
\cite{nature}.

In particular, for a value of
$E_{\rm VP}^{\rm(rel)}-E_{\rm VP}^{(0)}$ in the $2p_{1/2}-2s_{1/2}$ splitting,
the original results of the papers mentioned read $-0.0041$~meV
in~\cite{pachucki1}, $-0.0041$~meV in~\cite{borie} and $-0.002084$~meV
in~\cite{jentschura}.

The results obviously strongly disagree. We note that in principle
the quoted calculations treated the
$(Z\alpha)^4m(m/M)^2$ term differently. However, the later is smaller by a
factor of $m/M\sim 0.1$ and cannot be responsible for such a large
difference by factor of two for
$E_{\rm VP}^{\rm(rel)}-E_{\rm VP}^{(0)}$.

\begin{figure}[htpb]
\begin{center}
\resizebox{0.14\textwidth}{!} {\includegraphics{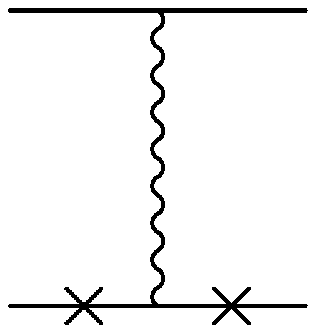}}
\resizebox{0.05\textwidth}{!} {\ }
\resizebox{0.14\textwidth}{!} {\includegraphics{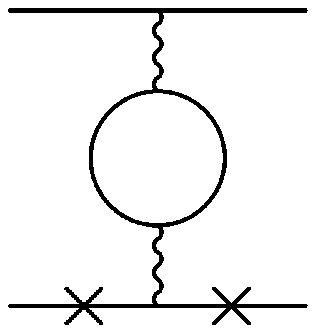}}
 \put(-145,-20){(a)}
 \put(-42,-20){(b)}
\end{center}
\caption{One-photon exchange diagram: for the pure photon exchange
(a) and for the eVP contributions (b).}
\label{f:1gamma}
\end{figure}

Once we consider two-body diagrams, we have to realize that in
principle there may be one-photon and two-photon contributions and
in principle the one-photon term (see. Fig.~\ref{f:1gamma}) includes
a static part (found at $k_0=0$) and a retardation part
(proportional to $k_0$ or $k_0^2$).

The partial results, such as the non-recoil one-photon contribution,
are not gauge invariant and only the sum of static one-photon,
retardation one-photon and two-photon contributions is gauge
invariant.

The dominant part in any reasonable gauge is due to static
one-photon exchange in order $\alpha(Z\alpha)^4m$. (In principle,
one can choose a ridiculous gauge with, e.g., a longitudinal part with a
parameter $\gg 1$. That would allow to obtain a `large' contribution
beyond the mentioned above terms, but that does not have much
physical sense.) The static one-photon exchange easily allows an
efficient evaluation both within the Breit-type and Grotch-type
approaches.

However, those terms have already been covered by a consideration of
the Dirac equation with the reduced mass. The purpose of this paper
(as well as of~\cite{pachucki1,borie,jentschura}) is to calculate
recoil corrections in order $\alpha(Z\alpha)^4m^2/M$ and
$\alpha(Z\alpha)^4m^3/M^2$ and we need to go beyond the leading
terms.

There is no `objective' separation between static one-photon,
retardation one-photon and two-photon contributions. E.g.,
considering the $(Z\alpha)^4m$ contribution (in all orders in $m/M$)
in different gauges, we have a kind of interplay of such terms.

Using the Coulomb gauge we find that for $(Z\alpha)^4m$ there is
neither a retardation correction nor a two-photon correction in order
of interest and thus the complete result in the $(Z\alpha)^4m$
contribution may be achieved through an application of either the Breit
equation (in all orders in $m/M$) or the Grotch equation for the
leading $m/M$ correction by applying the static one-photon kernel.

Now we return to the results from~\cite{pachucki1,borie,jentschura}
cited above. All of them are obtained by calculating the
static one-photon contributions. However, we demonstrate below that
different gauges were in fact applied and only in one of them the
retardation and two-photon contributions vanish while in the other
they do not.

\begin{figure}[htbp]
\begin{center}
\begin{picture}(0,0)
  \put(-20,30){\large (a)}
  \put(-20,-50){\large (b)}
  \end{picture}
  \resizebox{0.40\textwidth}{!} {
  \includegraphics{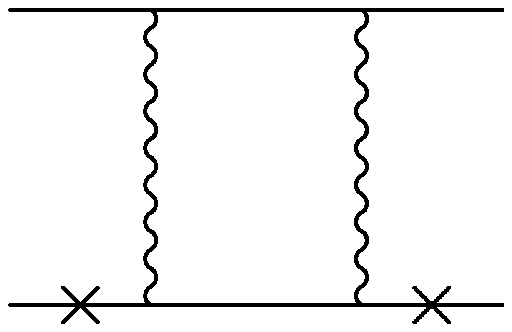} \resizebox{0.03\textwidth}{!} {\ } \includegraphics{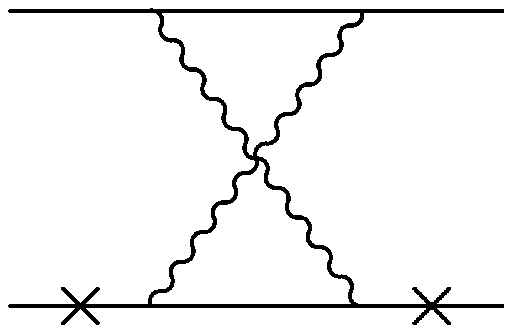}
  }
  \\[3ex]
  \resizebox{0.40\textwidth}{!} {
  \includegraphics{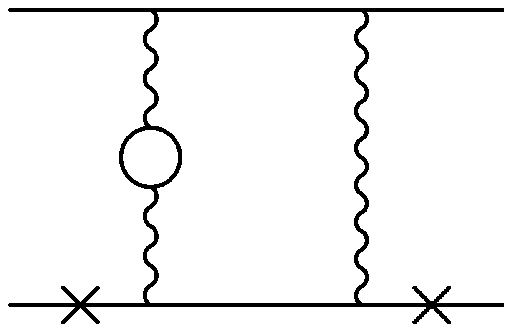} \resizebox{0.03\textwidth}{!} {\ } \includegraphics{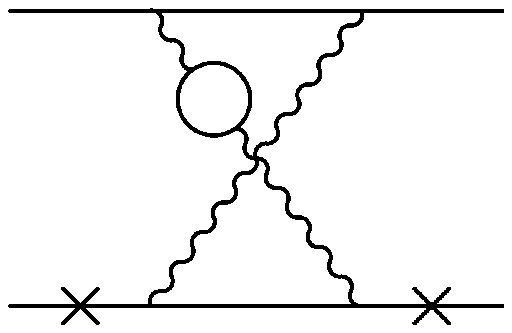}
  }
\end{center}
\caption{Two-photon exchange diagram: for the pure photon exchange
(a) and for the eVP contributions (b).}
\label{f:2gamma}
\end{figure}
We also note that the physical derivation of the Breit-type and
Grotch-type equations should actually start from a two-body
Bethe-Salpeter equation, next the latter should be reduced to an
effective one-body Dirac or two-body Schr\"odinger equation and
afterwards contributions of retardation and two-photon effects
should be estimated. The crucial part is to reduce the contribution
of interest to an one-photon contribution to the kernel of the
effective equation (see, e.g.,~\cite{ede}). A further mathematical
transformation of the static one-photon contribution is rather a
technical issue.

A naive two-photon contribution, free of eVP, (see
Fig.~\ref{f:2gamma}{\em a}) has infrared divergencies/singularities
which indicate that it includes in fact a correction of lower order,
namely the Coulomb correction $(Z\alpha)^2m$. (A divergence appears
once we neglect the atomic energy and momentum and consider the
related diagrams as free scattering diagrams, otherwise we should
speak about singularities.)

The derivation through various effective approaches generates
subtracted two-photon graphs (see, e.g.,~\cite{ede}). That is not a
trivial issue.

For example, if we choose the external-field approach for the first
approximation, then we should somehow `upgrade' $(Z\alpha)^2m$ up to
$(Z\alpha)^2m_R$. The missing $(Z\alpha)^2m^2/M$ term comes from a
two-photon contribution. This example shows that a rearrangement of
the diagrams applying a proper subtraction is crucially
important.

\begin{figure}[htpb]
\begin{center}
  \resizebox{0.45\textwidth}{!} {
  \includegraphics{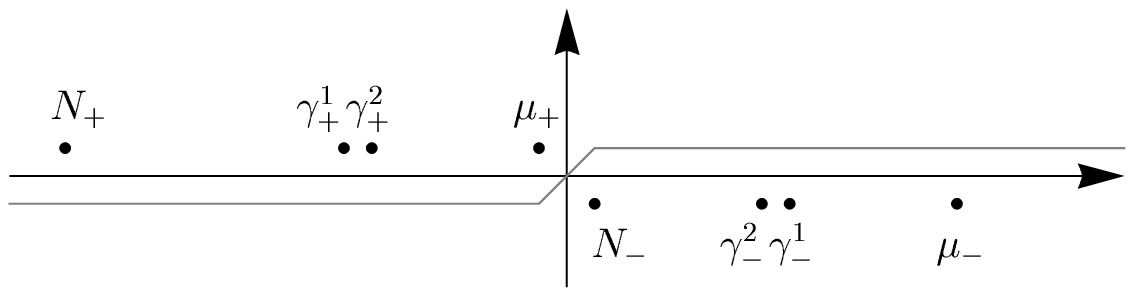}}
  \resizebox{0.45\textwidth}{!}{\includegraphics{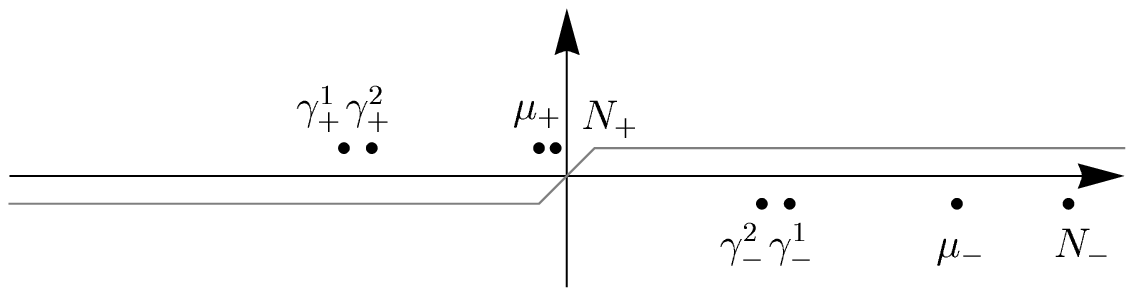}
  }
\end{center}
\caption{Poles for the unsubtracted two-photon exchange diagrams
(see Fig.~\ref{f:2gamma}). The upper plot is for the ladder diagrams
(right graphs in Fig.~\ref{f:2gamma}), while the lower one is for
the cross diagrams
(left graphs). For muonic hydrogen with $\kappa\sim 1$ the pole
structure for free diagrams and for eVP contributions (see Fig.
\ref{f:2gamma} {\em a\/} and {\em b\/}, respectively) is similar.
Here, $N$ stands for nuclear poles, $\gamma$ is for photonic poles and
$\mu$ is for muonic poles.}
\label{f:2poles}
\end{figure}

The $(Z\alpha)^2m^2/M$ correction is a result of the calculation of the
nuclear-pole contribution of the two-photon box diagram. The
effective-Dirac-equation approach suggests a complete subtraction of
the pole of the heavy particle \cite{ede} (see Fig.~\ref{f:2poles} for the pole
structure of unsubtracted two-photon diagrams of
Fig~\ref{f:2gamma}{\em a\/}; the pole of interest is denoted as
$N_-$).

Once such a pole is subtracted, we see that the one-photon
contribution is the only contribution for $(Z\alpha)^2m$ terms in
all orders of $m/M$ as well as the non-recoil part of the $(Z\alpha)^4m$
contribution. However, a recoil part of the $(Z\alpha)^4m$ term
arises in different gauges in different ways. The situation for the
eVP (see Fig~\ref{f:2gamma}{\em b}) is quite similar and the pole
structure is also similar.

Because of this similarity we briefly recall the situation with the
$(Z\alpha)^4m$ terms (in all orders in $m/M$) in different gauges.

The electromagnetic interaction of two particles in different gauges
is determined by the shape of the photon propagator $D_{\mu\nu}(k)$.
The term $(Z\alpha)^2m$ originates from the $D_{00}(k)$ component, while
$(Z\alpha)^4m$ corrections come from all components of
$D_{\mu\nu}(k)$.

One can immediately see that in any covariant gauge (in contrast to
the Coulomb gauge) the $D_{00}$ component depends on the energy
transfer $k_0$ and the retardation one-photon exchange produces a
correction of order $(Z\alpha)^4m(m/M)^2$. In the case of non-zero
values for the $D_{i0}$ components certain terms of order
$(Z\alpha)^4(m^2/M)$ can also appear from related one-photon
contributions.

The static one-photon contribution obviously differs in
different gauges and, after taking into account the retardation
terms, the one-photon contributions still differ. One should take
into account the two-photon diagrams to obtain complete
$(Z\alpha)^4(m^2/M)$ and $(Z\alpha)^4m(m/M)^2$ contributions, which
are indeed gauge invariant.

It is easy to estimate a `nominal' order of a two-photon diagram
suggesting that it converges if we neglect all the atomic effects.
The order is $(Z\alpha)^5m^2/M$. To obtain a lower order in
$Z\alpha$, such as $(Z\alpha)^4m^2/M$, we have to find terms
divergent at low momentum.

After the heavy-pole contribution is subtracted completely, the only
potentially divergent contributions are due to photon poles, if we
close the contour in the lower half-plane (see Fig.~\ref{f:2poles}).

The two-photon contribution in the Coulomb gauge has only a
logarithmic divergence, which cannot change the fact that two-photon
effects contribute in order $(Z\alpha)^5m^2/M$.

If we consider another gauge, such as the Feynman or Landau gauge,
the $(Z\alpha)^4m^2/M$ terms do appear from the photonic pole
contributions. We can see that such poles are important only in
diagrams with $D_{00}$ components for both photons, or with
$D_{i0}(q)$ contributions.

In the Coulomb gauge $D_{i0}=0$ and the $D_{00}$ component of the photon
propagator does not produce a pole and technically that is why the
two-photon exchange in the Coulomb gauge does not produce any
$(Z\alpha)^4(m^2/M)$ and $(Z\alpha)^4m(m/M)^2$ contributions.

Once the heavy-particle pole is subtracted, the one-photon
contribution is the only contribution for the $(Z\alpha)^4m$ terms in
all orders of $m/M$ in the Coulomb gauge.

\section{${\rm\textbf{eVP}}$-corrected photon propagator in various gauges}

Taking into account the transverse structure of the photon
self-energy caused by the vacuum polarization tensor
\begin{equation}
{\cal P}_{\mu\nu}(k)=\left(g_{\mu\nu}-\frac{k_\mu
k_\nu}{k^2}\right)\,{\cal P}(k^2)\;,
\end{equation}
one can derive in the Landau gauge
\begin{equation}
D_{\mu\nu}^{\rm LeVP}(k)=\frac{1}{k^2}\left[g_{\mu\nu}-\frac{k_\mu
k_\nu}{k^2}\right]\left[1+\frac{{\cal P}(k^2)}{k^2}\right]\;,
\end{equation}
or
\begin{equation}\label{DL1P}
D_{\mu\nu}^{\rm eVP}(k)=\frac{g_{\mu\nu}}{k^2}\left[1
+\frac{{\cal P}(k^2)}{k^2}\right]\;,
\end{equation}
which differs only by longitudinal terms resulting from an
obvious gauge transformation.

Here, for the eVP we apply Schwinger's parametrization
\begin{equation}\label{defP}
{\cal
P}(k^2)=\frac{\alpha}{\pi}\,\int_0^1{dv}\rho_e(v)\frac{k^4}{k^2-\lambda^2}\;,
\end{equation}
where
\begin{eqnarray}
\rho_e(v)&=&\frac{v^2(1-v^2/3)}{1-v^2}\\
\lambda^2&=&\frac{4m_e^2}{1-v^2} \;.
\end{eqnarray}

Indeed, neither the Landau nor Feynman gauge is well-suited for the
bound-state calculations and it is helpful first to perform a gauge
transformation on the photon propagator (\ref{DL1P}) to reach a more
suitable gauge.

The photon propagator with an eVP correction in an arbitrary gauge can
be presented in the form
\begin{equation}
D_{\mu\nu}(k)=D_{\mu\nu}^{\rm eVP}(k)+\chi_\mu k_\nu + k_\mu
\chi_\nu,
\end{equation}
where $\chi=\chi(k)$ is an arbitrary function of $k$.

We expect that in recent calculations~\cite{pachucki1,borie,jentschura}
of the relativistic recoil
corrections to eVP different gauges were used. In this paper we
consider two choices of the gauge function $\chi(k)$ and two
gauges.

The gauge function can be presented as an expansion in powers of $\alpha$
\begin{equation}
\chi=\chi^{(0)}+\frac{\alpha}{\pi}\chi^{(1)}\;.
\end{equation}
For the $\chi^{(0)}$ we chose the transformation which is to produce
the Coulomb gauge (for the free propagator), while for $\chi^{(1)}$
we consider two options, which are presented below.

The first choice is
\begin{eqnarray}\label{choice1}
\chi^{(1)}_0&=&-\int_0^1{dv}\rho_e \frac{k_0}{2(k^2-\lambda^2)({\bf k}^2+\lambda^2)}\;,\nonumber\\
\chi^{(1)}_i&=&\int_0^1{dv}\rho_e \frac{k_i}{2(k^2-\lambda^2)({\bf k}^2+\lambda^2)}\;,\nonumber\\
D_{00}&=&-\frac{\alpha}{\pi}\int_0^1{dv}\rho_e \frac{1}{{\bf k}^2+\lambda^2}\;, \nonumber\\
D_{i0}&=&0\;,\nonumber\\
D_{ij}&=&-\frac{\alpha}{\pi}\,\int_0^1{dv}\rho_e \frac{1}{k^2-\lambda^2}\nonumber\\
&~& \times \left(\delta_{ij}-\frac{k_ik_j}{({\bf
k}^2+\lambda^2)}\right)\;.
\end{eqnarray}
Let us refer to it as the C1eVP gauge.

A second possibility we consider here is
\begin{eqnarray}\label{choice2}
\chi^{(1)}_0&=&-\int_0^1{dv}\rho_e \frac{k_0}{2(k^2-\lambda^2){\bf k^2}}\;,\nonumber\\
\chi^{(1)}_i&=&\int_0^1{dv}\rho_e \frac{k_i}{2(k^2-\lambda^2){\bf k^2}}\;,\nonumber\\
D_{00}&=&-\frac{\alpha}{\pi}\,\int_0^1{dv}\rho_e \frac{1}{k^2-\lambda^2}\,\frac{k^2}{{\bf k^2}}\;, \nonumber\\
D_{i0}&=&0\;,\nonumber\\
D_{ij}&=&-\frac{\alpha}{\pi}\,\int_0^1{dv}\rho_e \frac{1}{k^2-\lambda^2}\nonumber\\
&~& \times\left(\delta_{ij}-\frac{k_ik_j}{{\bf k^2}}\right)\;.
\end{eqnarray}
We refer to it as the C2eVP gauge.

In the static regime (i.e., $k_0=0$) the choice in Eqs.~(\ref{choice1})
reproduces the potential applied in~\cite{pachucki1,jentschura},
while the choice in Eqs.~(\ref{choice2}) leads to a potential considered
in~\cite{borie}.

We note that the gauge (\ref{choice1}) is similar to the Coulomb
gauge in a sense that the $D_{00}$ has no dependence on $k_0$
and there is no $D_{i0}$ component. That means that the static
one-photon contribution should produce a complete result. There are
two contradictory results for the $2p-2s$ Lamb splitting in
literature~\cite{pachucki1,jentschura} and we confirm the
latter result. Our result for muonic hydrogen is%
\footnote{Unless otherwise stated, the uncertainty is equal to unity
in the last presented digit.}
\begin{eqnarray}
\Delta E_{\rm eVP}(2p_{1/2}-2s_{1/2})=0.0187589\ {\rm meV}\,.
\end{eqnarray}

Considering the gauge (\ref{choice2}) we note that, while $D_{i0}=0$,
the $D_{00}$ component of the propagator depends on $k_0$ through
the eVP tensor in Eq.~(\ref{defP}) and one not only has to check the
static one-photon term, but also calculate the one-photon
retardation part and examine the photonic pole contributions for
the two-photon diagrams (see Fig.~\ref{f:2gamma}{\em b}).

To check the consistency of the result obtained and to compare with
other existing calculations, we perform below four separate
calculations applying either a C1eVP gauge or C2eVP gauge within
either the Breit-type or Grotch-type approach.

\section{Calculation in the ${\rm \textbf{C1eVP}}$ gauge~(\ref{choice1})}

Now, let us perform the calculations in the C1eVP gauge
(\ref{choice1}), which because of lack of retardation and two-photon
contributions in order $\alpha(Z\alpha)^4m$ (in all orders in
$(Z\alpha)$) should produce a correct result in an easier way.

\subsection{Breit-type calculation}

We start with a Breit-type calculations.

First, we note that as it is well known (see, e.g.,~\cite{breit1})
the energy of hydrogenic levels without eVP can be found by
considering a non-relativistic Schr\"odinger-type equation with a
Hamiltonian
\begin{eqnarray}
H&=&H_0+H_1\;,\nonumber
\end{eqnarray}
\begin{eqnarray}
H_0&=&\frac{{\bf p}^2}{2m_R}+V_C=\frac{{\bf p}^2}{2m_R}-\frac{Z\alpha}{r}\;,\nonumber
\end{eqnarray}
\begin{eqnarray}\label{breit0}
H_1&=&-\frac{{\bf p}^4}{8}\left(\frac{1}{m^3}+\frac{1}{M^3}\right)+\frac{Z\alpha \pi}{2}\left(\frac{1}{m^2}+\frac{1}{M^2}\right)\delta^3(r) \nonumber\\
&-&\frac{Z\alpha}{2mM\,r}\left({\bf p}^2+\frac{({\bf r}\cdot({\bf r}\cdot {\bf p}){\bf p})}{r^2}\right)\nonumber\\
&+&\frac{Z\alpha}{r^3}\left(\frac{1}{4m^2}+\frac{1}{2mM}\right)(\mbox{\boldmath $\sigma$}\cdot[{\bf r}\times{\bf p}])\;.
\end{eqnarray}
Here and in further considerations we ignore the nuclear spin
terms, assuming that the results are for the center of gravity of the
related hyperfine multiplet.

The result is obtained in the Coulomb gauge which is consistent with
both C1eVP and C2eVP gauge we are to consider. Here the first
term is responsible for the $(Z\alpha)^2m$ terms and the second
produces $(Z\alpha)^4m$ contributions (in all orders in $m/M$). To
find the former one has to solve the related Schr\"odinger equation
(let us denote the energy and wave functions as $E_0$ and $\Psi_0$)
and to find the latter one has to find a matrix element of $H_1$
over the Schr\"odinger-equation wave functions $\Psi_0$.

Applying the C1eVP gauge we find the additional terms in momentum
space which are necessary to
take into account eVP effects
\begin{eqnarray}\label{breit1imp}
  H^{\rm eVP}&=&H_0^{\rm eVP}+H_1^{\rm eVP}\;,\nonumber\\
  H^{\rm eVP}_0&=&V_U({\bf k})=-4\alpha (Z\alpha) \int\limits_0^1dv \, \rho_e \frac{1}{{\bf k}^2+\lambda^2} \;, \nonumber\\
  H_1^{\rm eVP}&=&-4\alpha (Z\alpha) \int\limits_0^1dv \, \rho_e \frac{1}{{\bf k}^2+\lambda^2}\nonumber\\
  &\times&\Biggl\{\left(-\frac{{\bf k}^2}{2}+i\mbox{\boldmath $\sigma$}\cdot[{\bf p}_i\times{\bf p}_f]\right)\left(\frac{1}{4m^2}+\frac{1}{4M^2}\right) \nonumber\\
  &+&\frac{1}{4Mm}\Biggl(({\bf p}_i+{\bf p}_f)\cdot\left(i[\mbox{\boldmath $\sigma$}\times{\bf k}]+{\bf p}_i+{\bf k}_f\right)\nonumber\\
  & &\qquad\qquad\qquad\qquad-\frac{\left({\bf p}_i^2-{\bf p}_f^2\right)^2}{({\bf
  k}^2+\lambda^2)}\Biggr)\Biggr\}\;.
\end{eqnarray}

The related expressions in coordinate space are
\begin{eqnarray}\label{breit1}
H^{\rm eVP}_0&=&V_U(r)=-\frac{\alpha}{\pi} (Z\alpha) \int\limits_0^1dv \, \rho_e  \frac{e^{-rs_e}}{r} \;, \nonumber\\
H^{\rm eVP}_1&=&\left(\frac{1}{8m^2}+\frac{1}{8M^2}\right)\nabla^2 V_U \nonumber\\
 &+&\left(\frac{1}{4m^2}+\frac{1}{2mM}\right)\frac{V_U'}{r}{\bf L}\cdot{\mbox{\boldmath $\sigma$}}
  \nonumber\\
 &+&\frac{1}{2mM}\nabla^2\left[V_U-\frac{1}{4}(rV_U)'\right] \nonumber \\
 &+&\frac{1}{2mM}\biggl[\frac{V_U'}{r}L^2+\frac{{\bf p}^2}{2}(V_U-rV_U') \nonumber\\
 &&\phantom{\frac{1}{2mM}\biggl[} +(V_U-rV_U')\frac{{\bf p}^2}{2}\biggr]\;.
\end{eqnarray}
The Hamiltonian $H^{\rm eVP}$ completely agrees with the one appearing in
\cite{pachucki1}.

Again, the first term is responsible for the $\alpha(Z\alpha)^2m$
terms and the second one produces $\alpha(Z\alpha)^4m$ contributions
(in all orders in $m/M$). However, now the procedure is somewhat
different. We are interested only in the first order in $\alpha$
results and thus we can consider both $H^{\rm eVP}_0$ and $H^{\rm
eVP}_1$ as perturbations.

To find the leading non-relativistic terms (see (\ref{u:lead})), we
have to calculate $\langle \Psi_0\vert H^{\rm eVP}_0\vert\Psi_0
\rangle$, however, a similar contribution of higher order,
$\langle \Psi_0\vert H^{\rm eVP}_1\vert\Psi_0 \rangle$, gives only a
part of the $\alpha(Z\alpha)^4m$ result. The other part results
from second-order perturbation theory on the Schr\"odinger equation
and it is of the form (see, e.g., \cite{pachucki1})
\begin{equation}\label{2nd:pt}
2 \langle \Psi_0\vert H^{\rm eVP}_0
\frac{1}{\left(E-H_0\right)^\prime} H_1 \vert\Psi_0 \rangle\;,
\end{equation}
where the reduced Green function
$\frac{1}{\left(E-H_0\right)^\prime}$ is applied. In our
calculations for the non-relativistic reduced Coulomb Green function
we used its presentation in terms of smaller and larger radii
(cf.~\cite{pachucki2}).

The wave function $\Psi_0$ is expressed in terms of the reduced mass
$m_R$ while the dependence on $m$ and $M$ is due to apparent factors
in Eqs.~(\ref{breit0}) and (\ref{breit1}) which allows to express results
of our calculation of various matrix elements in terms of $C_0$,
$C_1$ and $C_2$ as defined in Eq.~(\ref{recoilsplit}). That is helpful
for further comparison with other calculations.

Indeed, the $C_0$ results reproduce the values (\ref{allc0}),
as they should, while for other coefficients we find
\begin{eqnarray}\label{breit1c1}
C_1(1s)&=& 0.18153\dots\;, \nonumber\\
C_1(2s)&=& 0.038089\dots \;,\nonumber\\
C_1(2p_{1/2})&=& 0.00090127\dots\;, \nonumber\\
C_1(2p_{3/2})&=& 0.00090127\dots
\end{eqnarray}
and
\begin{eqnarray}\label{breit1c2}
C_2(1s)&=& -0.3631\dots\;, \nonumber\\
C_2(2s)&=& -0.07618\dots \;,\nonumber\\
C_2(2p_{1/2})&=& 0.003542\dots\;, \nonumber\\
C_2(2p_{3/2})&=& -0.004475\dots\;.
\end{eqnarray}
As we mentioned, the result for $\Delta E_{\rm eVP}(2p_{1/2}-2s_{1/2})$ is
consistent with the result of~\cite{jentschura}.

\subsection{Grotch-type calculation\label{s:G1}}

The Grotch type of approach includes a few operations (see~\cite{grotch}
for more detail). First, we have to present the two-body
wave function as a product of a free-spinor for the nucleus and
a four-component muon wave function. The potential is averaged over
the nuclear spinor.

The approach allows to reproduce the Dirac equation (with the
reduced mass) and obtain the leading recoil correction in order
$m/M$. After we neglect terms of higher order in $m/M$, we arrive at
an equation
 \begin{equation}
K \psi_0
  =E \psi_0 \,,
 \end{equation}
 where
\begin{eqnarray}\label{Grotch-eqn}
K  &=&
  \mbox{\boldmath$\alpha$} \cdot {\mathbf p}+\beta m+\frac{\mathbf{p}^2}{2M}+ V \\ \nonumber &&
    + \frac{1}{2M} \left\{ \mbox{\boldmath$\alpha$} \cdot {\mathbf p}, V \right\}
    + \frac{1}{4M}\left[\mbox{\boldmath$\alpha$} \cdot {\mathbf p},[\mathbf{p}^2, W]\right]
   \,,
\end{eqnarray}
and
\[W({\bf q}) =-\frac{2V({\bf q})}{{\bf q}^2}\;.\]

To find a solution it is helpful, following~\cite{grotch}, to
introduce an auxiliary Hamiltonian
 \begin{equation}
   K_1 = \mbox{\boldmath$\alpha$} \cdot {\mathbf p} + \beta m + V \frac{1-\beta
   m/M}{1-(m/M)^2}\,.
 \end{equation}
we note that
 \begin{equation}
 K=K_0+\Delta K  + {\cal O} \left( \frac{m^3}{M^2} (Z\alpha)^4 \right) \,,
 \end{equation}
where
\begin{equation}
 K_0=K_1+\frac{K_1^2-m^2}{2M}+\frac{1}{4M} [K_1,[\mathbf{p}^2,W]]\,,
\end{equation}
and
\begin{equation}
 \Delta K =-\frac{V^2}{2M}-\frac{1}{4M} [V,[\mathbf{p}^2,W]]\;.
\end{equation}

As it is known~\cite{grotch} for the case of pure Coulomb potential,
$\Delta K =0$.
Let us for the moment neglect $\Delta K$ for an
arbitrary potential and look for a wave function of the
form
\begin{equation}
  \psi_0 = N\, \left( 1-\frac{1}{4M} [ \mathbf{p}^2, W] \right)
  ( 1 + \beta \mu ) \widetilde\psi  \,.
\end{equation}
Since the Grotch-type approach does not control the $(m/M)^2$ terms,
below we expand in $m/M$ and neglect higher-order terms everywhere
where it is possible. The
results of such an expansion are denoted with
`$\simeq$'.

In particular, we find for the normalization constant~$N$
\begin{equation}
N = \frac{1}{\sqrt{1+2\mu\,\widetilde{E}/\widetilde{m}+\mu^2}}\simeq
1-\frac{\widetilde{E}}{2M}\,,\nonumber
\end{equation}
where the involved parameters are defined below.

The final Grotch-type equation takes the form of an effective
Dirac equation
 \begin{equation}\label{eqn-tilde}
 \left(\mbox{\boldmath$\alpha$} \cdot {\mathbf p} + \beta \widetilde{m}
 +\widetilde{V}\right)\widetilde\psi
 \simeq \widetilde{E}\widetilde\psi\,,
 \end{equation}
  \begin{equation}
 \widetilde{V}=\frac{V}{\sqrt{1-\frac{m^2}{M^2}}}\simeq V\,.
 \end{equation}
Solutions of the Dirac equation (\ref{eqn-tilde}),
$\widetilde\psi$ and $\widetilde{E}$, can be found, since the
final equation takes form of a Dirac equation with potential $V$
and various effective parameters. The identities for
$\widetilde\psi$ and $\widetilde{E}$ are of the same functional form
as for a solution of a Dirac equation, but they express the wave
functions and energy in terms of effective parameters defined as
\begin{eqnarray}
  \mu &=&  \frac{M}{m} \left( 1-\sqrt{1-\frac{m^2}{M^2}} \right)\simeq \frac{m}{2M}\,,\nonumber\\
      E_0&=&E_1 + \frac{E_1^2-m^2}{2M}\,,\nonumber\\
\widetilde{E}&=&\frac{E_1-m^2/M}{\sqrt{1-m^2/M^2}}\simeq E_1-m^2/M\,,\nonumber\\
\widetilde m &=& \frac{m\left( 1-\frac{E_1}{M}
\right)}{\sqrt{1-\frac{m^2}{M^2}}}\simeq m\left( 1-\frac{E_1}{M}
\right)\,,
\end{eqnarray}

One can solve the equation (\ref{eqn-tilde}) as far as the solution of the
conventional Dirac equation is known for a potential $V$ with an
appropriate accuracy.

The final energy has a correction due to $\Delta K$, which was
neglected in order to obtain a solvable equation. The final
result for the energy is
 \begin{equation}
E \simeq E_0 + \langle \psi_0 \vert \Delta K \vert \psi_0\rangle\,.
 \end{equation}
We remind that $\Delta K \propto m/M$ and here we neglect all
$(m/M)^2$ corrections.

Now, one can introduce the potential. In the C2eVP gauge, because
the photon propagator is proportional to the free propagator in the
Coulomb gauge, the equation would take the same form as for the
Coulomb potential, but now with potential.
 \begin{equation}\label{pot:cu}
V=V_C+V_U\,.
 \end{equation}
However, in the case of the C1eVP gauge, we have to introduce a
certain correction, namely by redefining $W$,
\begin{eqnarray}
 V&=&V_C+V_U\,,\nonumber\\
 W&=&W_C+W_U\,,\nonumber\\
 W_C&=&-\frac{2V_C(q)}{q^2}\,,\nonumber\\
 W_U&=&8\alpha(Z\alpha) \int_0^1 dv \,
 \frac{\rho_e(v)}{(q^2+\lambda^2)^2}\;.
\end{eqnarray}

In both cases, we have to solve an effective Dirac equation
(\ref{Grotch-eqn}) with a potential (\ref{pot:cu}). The energy levels
for such a Dirac equation (with a reduced mass for the
particle) with $V=V_C$ are well known
\[
E_{\rm Coul}(nlj) = m + E^{\rm (NR)}_{\rm Coul} + E_{\rm Coul}^{(0)}
\]
\begin{equation}
\approx m - \frac{(Z\alpha)^2m_R}{2n^2} + \frac{(Z\alpha)^4m_R}{2n^3} \left( \frac{3}{4n}-\frac{1}{j+1/2} \right)
\end{equation}
and the linear in $\alpha$ correction due to eVP was calculated for
the Dirac wave functions as explained in Sect.~\ref{s:leading}
(see for details
Eqs.~(\ref{u:lead}), (\ref{recoilsplit}) and (\ref{allc0}))
\begin{equation}
E_{\rm VP} = E^{(\rm NR)}_{\rm VP} + E^{(0)}_{\rm VP}\;.
\end{equation}

Solving the above equations, one can arrive at (see \cite{elsewhere} for details)
\begin{eqnarray}
 E_0&\simeq &
 E_{\rm Coul}(nlj)+E_{\rm VP} \nonumber\\
  &-&\frac{E^{\rm (NR)}_{\rm Coul}}{M}\left[E_{\rm VP}^{\rm (NR)}
  +\kappa\frac{\partial}{\partial \kappa}E_{\rm VP}^{\rm (NR)}\right]
 \,.
\end{eqnarray}

We can now return to the $\Delta K$ term. The related correction in
the first order in eVP is a matrix element of
 \begin{eqnarray}
 \Delta K &\simeq&
  -\frac{1}{2M}\biggl(2V_U V_C+\frac{1}{2}[V_C,[\mathbf{p}^2,W_U]]\nonumber\\
  &&+\frac{1}{2}[V_U,[\mathbf{p}^2,W_C]]\biggr)\,.
 \end{eqnarray}
After a simple estimation of the operator, we find that it is
sufficient to calculate the matrix element using
Schr\"odinger-Coulomb wave functions, which are indeed well known.

More detail on the application of the Grotch-type approach to the eVP
contribution will be published elsewhere~\cite{elsewhere}.

Finally, we obtain the results
\begin{eqnarray}\label{grotch1c1}
C_1(1s)&=& 0.18153\dots\;, \nonumber\\
C_1(2s)&=& 0.038087\dots \;,\nonumber\\
C_1(2p_{1/2})&=&0.00090127\dots\;, \nonumber\\
C_1(2p_{3/2})&=& 0.00090127\dots\;,
\end{eqnarray}
which completely agree with the results (\ref{breit1c1}) of the Breit-type
calculation.

\section{Calculation in the ${\rm \textbf{C2eVP}}$ gauge~(\ref{choice2})}

For the C2eVP gauge (\ref{choice2}) we note that we have to
calculate a static one-photon exchange, retardation one-photon
contribution and two-photon contribution. Here, we first calculate
the static one-photon contribution applying the Breit-type and
Grotch-type techniques and afterwards we find the retardation
one-photon contribution and two-photon contribution as a
perturbation.

\subsection{Static one-photon exchange}

\subsubsection{Breit-type calculation}

The Breit-type Hamiltonian is somewhat different from
Eq.~(\ref{breit1imp}) and the addition is
\begin{eqnarray}\label{addition}
\frac{\alpha (Z\alpha)}{Mm}\int\limits_0^1dv \, \rho_e
\frac{\lambda^2\left({\bf p}_i^2-{\bf p}_f^2\right)^2}{{\bf
k}^2({\bf k}^2+\lambda^2)^2}
 \end{eqnarray}
which shifts the results for the static contribution.

We find for the static one-photon contribution%
\begin{eqnarray}\label{breit2c1}
C_1(1s)&=& 0.3981\dots\;, \nonumber\\
C_1(2s)&=& 0.06357\dots \;,\nonumber\\
C_1(2p_{1/2})&=&0.002845\dots\;, \nonumber\\
C_1(2p_{3/2})&=&0.002845\dots\;,
\end{eqnarray}
and
\begin{eqnarray}\label{breit2c2}
C_2(1s)&=& -0.7961\dots\;, \nonumber\\
C_2(2s)&=& -0.1271\dots \;,\nonumber\\
C_2(2p_{1/2})&=&-0.0003444\dots\;, \nonumber\\
C_2(2p_{3/2})&=&-0.008362\dots\;,
\end{eqnarray}
which indeed does not coincide with Eqs.~(\ref{breit1c1}) and
(\ref{breit1c2}), because such a contribution is not gauge
invariant.

\subsubsection{Grotch-type calculation}

A similar correction should be introduced into the Grotch-type
approach. As we already mentioned, since the C2eVP gauge is
proportional to the Coulomb gauge, we can use the same kind of
equation as for the Coulomb gauge with a potential
\[
V=V_C+V_U
\]
and the $W$ function defined within the same functional relation as
for the Coulomb potential, namely as
\[
W(q) =-\frac{2V(q)}{q^2}\;.
\]
The effective Dirac equation, which does not involve $W$, is indeed
the same as in C1eVP gauge, while the $\Delta K$ correction proportional to $W$
is different.

Proceeding similarly to described in Sect. \ref{s:G1} we arrive at
\begin{eqnarray}\label{grotch2c1}
C_1(1s)&=& 0.39818\dots\;, \nonumber\\
C_1(2s)&=& 0.063585\dots \;,\nonumber\\
C_1(2p_{1/2})&=&0.0028496\dots\;, \nonumber\\
C_1(2p_{3/2})&=&0.0028496\dots\;,
\end{eqnarray}
which is consistent with Eq.~(\ref{breit2c1}) and somewhat disagrees with~\cite{borie}.

In particular, our result for $E_{\rm VP}^{\rm(rel)}-E_{\rm
VP}^{(0)}$ for the $2p-2s$ splitting for the static contribution
is $-0.0042785$~meV which is to be compared with
$-0.0041$~meV in~\cite{borie}.

The relativistic recoil eVP correction in light muonic atoms was
calculated by Borie in~\cite{borie} in a way somewhat different from, but
consistent with our treatment here of the Grotch-type approach in
the gauge C2eVP.

In the recent paper~\cite{borie} some minor corrections to the
earlier papers~\cite{borie82} and \cite{borie05} were introduced.
Still, we failed to reproduce exactly the numerical results~\cite{borie} for
the correction neither by expressions given in~\cite{borie} nor by
those presented in~\cite{borie82}, where further details of the
calculation were given. Apparently, expressions of~\cite{borie}
still contain misprints.

What is more important, the results we obtained in this section also
disagree with the result~\cite{borie}. The departure grows systematically
between muonic hydrogen, deuterium and helium. (The results
for the latter are presented in Sect.~\ref{lslight} of our paper.)

Meanwhile, we have discovered that our expression for the energy
correction agrees with one presented in~\cite{borie82} (see
Eq.~(116); it is also reproduced in Appendix~A of~\cite{borie}). Our
numerical results can be reproduced if we modify the erroneous
expression for the term
$\left\langle\frac{Z\alpha}{3r^4}Q_4\right\rangle$ presented in
Appendix~A of \cite{borie}.

Thus, we conclude that the result~\cite{borie} for the relativistic
recoil correction in light muonic atoms is unfortunately both
incomplete because of lack of two-photon contributions and incorrect
(because even the partial calculation contains a numerical error).

\subsection{Retardation one-photon exchange}

The retardation one-photon contribution and an essential two-photon
contribution can be calculated directly as a perturbation,
since they are already related to effective potentials which are smaller
by a factor $(Z\alpha)^2$ than the non-relativistic
contributions.
It is also important that the second order perturbation term (see
(\ref{2nd:pt})) is the same as in C1eVP gauge. That is because the
free term of eVP propagator is the same for both gauges (which
determines $H_1$) and the static limit of the eVP term of the propagator
also does not change (which determines $H_0^{\rm eVP}$).
That says that only first-order perturbation theory terms are
essential, otherwise the second-order terms similar to
(\ref{2nd:pt}) would appear.

One can immediately find a related effective addition to the
Hamiltonian due to retardation
\begin{equation}\label{retard}
\frac{\alpha(Z\alpha)}{M^2}\int\limits_0^1dv \, \rho_e
\frac{\lambda^2\left({\bf p}_i^2-{\bf p}_f^2\right)^2}{{\bf k}^2({\bf k}^2+\lambda^2)^2}
\end{equation}
or in coordinate space
\begin{eqnarray}
H_{\rm retard}&=&
 \frac{\alpha(Z\alpha)}{4\pi\,M^2} \int_0^1 dv \rho(v) \\ \nonumber
 &\times&
 \left[\left\{{\bf p}^4,Q(r)\right\}-2{\bf p}^2Q(r)\,{\bf
 p}^2\right]\,,
\end{eqnarray}
where
\begin{equation}
   Q(r)=
   \frac{1}{\lambda^2r}-(2+rs_e)\frac{e^{-r s_e}}{2 \lambda^2 r}\,.
\end{equation}

The results of direct calculations of $C_1$ and $C_2$ in muonic
hydrogen are compiled in Tables~\ref{t:1c} and \ref{t:2c},
respectively.
The retardation effects under consideration are of
order $\alpha(Z\alpha)^4m(m/M)^2$ and contribute only to $C_2$.

\begin{table}
\begin{tabular}{|c|c|c|c|c|}
  \hline
  Contribution
  & $C_1(1s)$ & $C_1(2s)$ & $C_1(2p_{1/2})$ & $C_1(2p_{3/2})$ \\  \hline
  Static & $\phantom{-}0.39818$ & $\phantom{-}0.063585$ & $\phantom{-}0.0028496$ & $\phantom{-}0.0028496$ \\
  Retardation
  & $\phantom{-}0\phantom{.00000}$ & $\phantom{-}0\phantom{.000000}$ & $\phantom{-}0\phantom{.0000000}$ & $\phantom{-}0\phantom{.0000000}$ \\
  Two-photon
  & $-0.21654$ & $-0.025483$ & $-0.0019484$ & $-0.0019484$ \\
\hline
  Total & $\phantom{-}0.18164$ & $\phantom{-}0.038102$ & $\phantom{-}0.0009012$ & $\phantom{-}0.0009012$ \\
  \hline
\end{tabular}
\caption{$C_1$ coefficients in the C2eVP gauge.}\label{t:1c}
\end{table}

\begin{table}
\begin{tabular}{|c|c|c|c|c|}
  \hline
  Contribution
  & $C_2(1s)$ & $C_2(2s)$ & $C_2(2p_{1/2})$ & $C_2(2p_{3/2})$ \\  \hline
  Static       & $-0.7961$ & $-0.1271\phantom{0}$ & $-0.0003444$ & $-0.008362$ \\
  Retardation
  & $\phantom{-}0.2165$ & $\phantom{-}0.02548$ & $\phantom{-}0.001948\phantom{0}$ & $\phantom{-}0.001948$ \\
  Two-photon
  & $\phantom{-}0.2165$ & $\phantom{-}0.02548$ & $\phantom{-}0.001948\phantom{0}$ & $\phantom{-}0.001948$ \\
\hline
  Total & $-0.3631$ & $-0.07618$ & $\phantom{-}0.003552\phantom{0}$ & $-0.004465$ \\
  \hline
\end{tabular}
\caption{$C_2$ coefficients in the C2eVP gauge.}
\label{t:2c}
\end{table}

\subsection{Two-photon exchange}

For the two-photon exchange we have performed a calculation of the
photon pole contributions. One has to carefully consider splitting of
the contributions. We are interested in those which are singular at
low momentum. Actually the photon-pole contribution is divergent in a formal sense
at high momentum, but such a divergence is in order
$\alpha(Z\alpha)^5m^2/M$ and thus is of a higher order. One has to
separate properly the low-momentum and high-momentum contributions, and
after that only the low-momentum one is of interest. The results are
summarized in Tables~\ref{t:1c} and~\ref{t:2c},
respectively.

We note that the sum of the results in the C2eVP gauge produces a
result consistent with those in the C1eVP gauge. We may also find
an effective potential induced by a low-momentum contribution of the
two-photon kernel. It is of the form
\begin{eqnarray}\label{twopho1}
&-&\frac{\alpha(Z\alpha)^2}{2\pi M}\int\frac{d^3{\bf q}}{(2\pi)^3}\int\limits_0^1dv \, \rho_e\frac{\lambda^2}{({\bf q}^2+\lambda^2)^2}\nonumber\\
 &\times&\frac{4\pi}{{\bf q}^2}\frac{4\pi}{({\bf k}-{\bf q})^2}\left({\bf q}^2-{\bf k}\cdot{\bf q}\right)\,.
\end{eqnarray}
For the diagonal matrix elements we can replace the expression (\ref{twopho1})
by an effective potential of the form
\begin{equation}\label{twopho}
-\frac{\alpha(Z\alpha)}{Mm_R}\int\limits_0^1dv \,
\rho_e\lambda^2\frac{\left({\bf p}_i^2-{\bf p}_f^2\right)^2}{{\bf
k}^2({\bf  k}^2+\lambda^2)^2}\,.
\end{equation}

We see that at the end of the day the final effective Hamiltonians in
both gauges are the same. The addition in the C2eVP gauge
(\ref{addition}) for the static term is eventually cancelled out by
the retardation (\ref{retard}) and two-photon ({\ref{twopho}) terms:
\begin{eqnarray}\label{cancel}
0&=&\alpha(Z\alpha)\int\limits_0^1dv \, \rho_e\frac{\lambda^2\left({\bf
p}_i^2-{\bf p}_f^2\right)^2}{{\bf k}^2({\bf k}^2+\lambda^2)^2}\nonumber\\
&\times&
\left(
\frac{1}{Mm}+\frac{1}{M^2} -\frac{1}{Mm_R} \right)\,.
\end{eqnarray}

\section{Finite-nuclear-size corrections}

As we mentioned in the introduction, the definition of the nuclear
radius for different spin values can produce additional corrections
in order $(Z\alpha)^4m(m/M)^2$ and  $\alpha(Z\alpha)^4m(m/M)^2$. For
this reason we consider in this paper the finite-nuclear-size (FNS)
corrections. While the relation between the eVP recoil contributions
and such corrections is considered in the next section, here we
revisit the FNS contributions with inclusion of eVP.

Such contributions are known for a while~\cite{friarNS,pachucki2,EGS},
basically numerically~\cite{borie05}.
Here we present semi-analytic results.

We treat the FNS effects non-relativistically. In this case the FNS
correction is of the form
\begin{eqnarray}
\Delta E_{\rm FNS} &=& \frac {R^2_N}{6} \langle \Psi \vert
\nabla^2(V_C+V_U) \vert \Psi \rangle\nonumber\\
&=&
\Delta E_{\rm FNS}^{(0)}
+\Delta E_{\rm FNS}^{(1)}
+\Delta E_{\rm FNS}^{(1)}\;,
\end{eqnarray}
where the wave function is the result of the Schr\"odinger equation
with the reduced mass and potential $V_C+V_U$ and
$R_N$ is the rms nuclear charge radius.
The result is not vanishing only for the $s$ states.

The leading term for the splitting
\begin{equation}
 \Delta E_{\rm FNS}^{(0)}(2s) = \frac{(Z\alpha)^4}{12}\,(R_N^2 m_R^2)m_R
\end{equation}
is applied for a determination of the nuclear size radius from the experimental
Lamb shift value and it is important to calculate corrections to it.

The two corrections can be obtained in a way similar to a
non-relativistic calculation of the leading eVP correction to the
hyperfine structure (cf.~\cite{pachucki2,martynenko05,muhfs}).
The first term is
\begin{equation}
\Delta E_{\rm FNS}^{(1)} = \frac {2\pi Z\alpha}{3}R^2_N \left(\vert
\Psi(0)\vert^2- \vert \Psi_C(0)\vert^2 \right)\;.
\end{equation}

It is convenient to present the eVP correction in the form
\begin{equation}
\Delta E_{\rm FNS}^{(1)} = \Delta E_{\rm FNS}^{(0)} \; \frac{\vert
\Psi(0)\vert^2-\vert \Psi_C(0)\vert^2}{\vert \Psi_C(0)\vert^2} \;,
\end{equation}
where we express the correction in terms of the leading term
and the eVP correction to the wave function at the origin. The
latter was studied in~\cite{pi_1s,muhfs,soto,vp32}.

The other term is
\begin{equation}
\Delta E_{\rm FNS}^{(2)} = \frac {R^2_N}{6} \langle \Psi_C
\vert \nabla^2{V_U} \vert \Psi_C \rangle.
\end{equation}
It was also found in~\cite{pi_1s}:
\begin{equation}
\Delta E_{\rm FNS}^{(2)}(2s) = \frac{\alpha}{\pi}\, G_{2s}(\kappa/2)\,\Delta E_{\rm FNS}^{(0)}(2s)\;,
\end{equation}
where
\begin{eqnarray}
G_{2s}(z)&=&
-\frac{\pi}{3z^3}
+\frac{24-44z^2-29z^4+22z^6}{36z^2(1-z^2)^2}\nonumber\\
&+&\frac{8-20z^2+33z^4-20z^6+8z^8}{12z^3(1-z^2)^2}\,A(z)\,,\nonumber
\end{eqnarray}
where $A(z)$ is defined by Eq.~(\ref{Adef}).

The numerical results on the finite-nuclear-size corrections
for the $2s$ state in light muonic atoms are summarized in Table~\ref{t:ns}.
Note that for the Lamb shift the signs of the corrections are opposite.
The results are slightly different but rather consistent with ones presented
in~\cite{borie}.%
\footnote{We are grateful to E.~Borie who drew our attention
to inaccuracies in results for muonic helium.}

\begin{table}
\begin{tabular}{|c|c|c|c|c|}
  \hline
  Atom
  &$\Delta E_{\rm FNS}^{(0)}$
  &$\Delta E_{\rm FNS}^{(1)}$
  &$\Delta E_{\rm FNS}^{(2)}$
  &$\Delta E_{\rm FNS}$\\
  \hline
  $\mu$H      &5.1975   &0.0170 &0.0110 &5.2254\\
  $\mu$D      &6.0732   &0.0205 &0.0132 &6.1069\\
  $\mu^3$He   &102.52 &0.520\phantom{0} &0.323\phantom{0} &103.37\\
  $\mu^4$He   &105.32 &0.536\phantom{0} &0.333\phantom{0} &106.19\\
  \hline
\end{tabular}
\caption{The non-relativistic finite-nuclear-size corrections for the $2s$ state
in light muonic atoms (in meV$\cdot\,(R_N/{\rm fm})^2$).
Here, $\mu$He stands for muonic helium ions.}
\label{t:ns}
\end{table}

\section{Results for the Lamb shift in light
muonic atoms}\label{lslight}

To conclude a calculation of the relativistic recoil corrections, we
have to fix the definition of the nuclear charge radius,
which varies in literature. That is important for the
relativistic recoil corrections because with different definitions
of the nuclear charge radius a certain part of the
$(Z\alpha)^4(m/M)^2m$ and $\alpha(Z\alpha)^4(m/M)^2m$ terms may be incorporated in the
nuclear-finite-size term.

The nuclear spin takes different values for light muonic atoms,
namely, $I=1/2$ for a muonic hydrogen and muonic helium-3 ion, $I=0$ for
a muonic helium-4 ion and $I=1$ for muonic deuterium. The different
nuclear spin values are related to different effective two-body
Breit-type equations for structureless particles
(see, e.g.,~\cite{scalar,vector}).

To be consistent with the experimental determination, we use the same
definitions of the nuclear charge radius as applied
in~\cite{scalar,vector,vector1,friar,codata,nature}.

In this convention the Zitterbewegung term is present for
half-integer spin nuclei, and not present for the integer case.

The related calculation produces the results summarized in Table~\ref{t:ls}.

\begin{table}
\begin{tabular}{|c|c|c|}
  \hline
  Atom & $\Delta E_{\rm eVP}(2p_{1/2}-2s_{1/2})$ & $\Delta E_{\rm eVP}(2p_{3/2}-2p_{1/2})$ \\[1ex]  \hline
  $\mu$H       & 0.018759 & 0.0049638 \\
  $\mu$D       & 0.021781 & 0.0057361 \\
  $\mu^3$He    & 0.50934\phantom{0}  & 0.26920\phantom{00} \\
  $\mu^4$He    & 0.52110\phantom{0}  & 0.27502\phantom{00} \\
  \hline
\end{tabular}
\caption{Relativistic recoil corrections to the Lamb shift and fine structure in light muonic atoms (in~meV).}
\label{t:ls}
\end{table}

Our results for the Lamb shift in light muonic atoms agree with the
results of~\cite{jentschura} and disagree with the results
of~\cite{borie} and~\cite{martynenko07}.

For the fine structure, the results are also presented in Table~\ref{t:ls}.

The result has been obtained in the C1eVP gauge within Breit-type
calculations and to control them we also performed a Grotch-type
calculation in the same gauge. They are in perfect agreement with each
other.

\section{Summary}

Concluding, relativistic recoil contributions in orders
$\alpha(Z\alpha)^4m(m/M)$ and $\alpha(Z\alpha)^4m(m/M)^2$ to the
Lamb shift in muonic hydrogen were revisited. The results published
previously by various authors and obtained by different methods are
inconsistent. In particular, result of the Breit-type calculation
in~\cite{jentschura} is twice smaller than the related result from the
Grotch-type evaluation in~\cite{borie}. The value of discrepancy,
0.002~meV, is comparable with the experimental uncertainty of
0.003~meV~\cite{nature}.

We perform here an evaluation in both approaches and find that the
discrepancy between~\cite{jentschura} and~\cite{borie}
is caused by the fact
that both calculated the same value, namely the static one-photon
exchange, which is not gauge invariant. They applied different
gauges. The gauge invariant value is a sum of the static term and
two other contributions, which are the retardation correction and
two-photon contribution. While they are absent for the calculation
of the $\alpha(Z\alpha)^4m(m/M)$ term in one gauge, they are not
vanishing in the other.

Once such contributions are taken into account and a relatively
small numerical error in calculation~\cite{borie} is fixed, we find
perfect agreement between two approaches. Our results agree with
those of~\cite{jentschura}.

We also consider other light atoms and perform calculations for the
Lamb shift muonic deuterium and two isotopes of muonic helium. For a
muonic helium-4 ion we agree with~\cite{jentschura} and disagree with
\cite{martynenko07}.

For control purposes we have also performed calculations of the fine
structure in order $\alpha(Z\alpha)^4m(m/M)$ and
$\alpha(Z\alpha)^4m(m/M)^2$. The result for the latter can be
completely restored from the result from Dirac equation
(see, e.g.,~\cite{rel2}). Our result for muonic hydrogen is in agreement
with~\cite{fs_martynenko}.

\section*{Acknowledgments}

This work was supported in part by DFG (grant GZ: HA 1457/7-1), RFBR
(under grant \# 11-02-91343) and Dynasty foundation. The authors are
thankful to E.~Borie for constructive remarks to the manuscript. VGI
and EYK are grateful to the Max-Planck-Institut f\"ur Quantenoptik
for its warm hospitality. SGK is also grateful to UNSW for their
hospitality and Gordon Godfrey foundation for their fellowship.

\end{document}